\documentclass[epj]{webofc}
\usepackage[utf8]{inputenc}
\usepackage[varg]{txfonts} 
\usepackage{booktabs}
\usepackage{xcolor}
\usepackage{braket}
\definecolor{darkred}{rgb}{0.4,0.0,0.0}
\definecolor{darkgreen}{rgb}{0.0,0.4,0.0}
\definecolor{darkblue}{rgb}{0.0,0.0,0.4}
\usepackage[bookmarks,linktocpage,colorlinks,
  linkcolor = darkred,
  urlcolor  = darkblue,
  citecolor = darkgreen]{hyperref}
\usepackage{amsmath}
\usepackage{bm}
\usepackage{dsfont}
\usepackage{caption}
\usepackage{nicefrac}
\usepackage{braket}
\usepackage{float}
\usepackage{subfigure}
\wocname{EPJ Web of Conferences}
\woctitle{Lattice2017}

\usepackage[normalem]{ulem}
\newcommand\xoutpars[1]{\let\helpcmd\xout\parhelp#1\par\relax\relax}
\newcommand\soutpars[1]{\let\helpcmd\sout\parhelp#1\par\relax\relax}
\long\def\parhelp#1\par#2\relax{%
  \helpcmd{#1}\ifx\relax#2\else\par\parhelp#2\relax\fi%
}

\begin{document}
\selectlanguage{english}
\title{%
  A tmQCD mixed-action approach to flavour physics
}
\author{%
  \firstname{Gregorio} \lastname{Herdo\'{\i}za}\inst{1,2} 
  \and
  \firstname{Carlos} \lastname{Pena}\inst{1,2} 
  \and
  \firstname{David}  \lastname{Preti}\inst{1} 
  \and
  \firstname{José Ángel}
  \lastname{Romero}\inst{1}\fnsep\thanks{Speaker, \email{ja.romero@csic.es}}
  \and\\
  \firstname{Javier}
  \lastname{Ugarrio}\inst{1,2}\fnsep\thanks{Speaker, \email{javier.ugarrio@uam.es}}
}
\institute{
  Instituto de F\'{\i}sica Te\'orica UAM-CSIC, c/ Nicol\'as Cabrera 13-15, Universidad Autónoma de Madrid, \\ E-28049 Madrid, Spain
  \and
  Department of Theoretical Physics, Universidad Autónoma de Madrid, E-28049 Madrid, Spain
}
\abstract{ We discuss a mixed-action approach in which sea quarks are
  regularised using non-perturbatively ${\rm O}(a)$ improved Wilson
  fermions, while a fully-twisted tmQCD action is used for valence
  quarks. In this setup, automatic ${\rm O}(a)$ improvement is
  preserved for valence observables, apart from small residual ${\rm
    O}(a)$ effects from the sea. A strategy for matching sea and
  valence is set up, and carried out for $N_\mathrm{f}=2+1$ CLS
  ensembles with open boundary conditions at several simulation
  points. The scaling of basic light-quark observables such as the
  pseudoscalar meson decay constant is studied, as well as the isospin
  splitting of pseudoscalar meson masses. }
\begin{flushright}
  IFT-UAM/CSIC-17-109 \\ FTUAM-17-26
\end{flushright}
\maketitle
\section{Introduction}\label{intro}

Flavour physics is one of the most promising sectors for the search of
physics beyond the Standard Model (SM).  In particular, processes
involving heavy quarks play a crucial role in the ongoing searches of
New Physics phenomena.  Current (LHCb) and upcoming (Belle II)
experiments will considerably improve the amount and precision of
available results. At this stage, the verification of the unitarity of
the CKM matrix~\cite{Cabbibo:1964zsa, Kobayashi:1973fv} as well as the
study of rare decays, appear as meaningful ways to examine the
compatibility between the SM predictions and the experimental
measurements.

Lattice QCD (LQCD) computations in the heavy quark sector are
essential to achieve reliable determinations of SM observables.  Our
main target is to increase the precision of LQCD calculations of
hadronic matrix elements necessary for the determination of CKM matrix
elements and of other quantities of phenomenological interest in the
charm sector. While the accuracy of LQCD computations of decay
constants relevant for $D_\mathrm{(s)}$-meson decays are in the same
ballpark as the current experimental precision of leptonic channels,
this is still not the case for other charmed mesons, nor in
semileptonic decays.  In particular, an accurate computation of the
form factors involved in $D \to \pi \ell \nu_{\ell}$ and $D \to K \ell
\nu_{\ell}$ decays is essential to making the most of the new
experimental results.

We have developed a setup that aims at optimising the control of
systematic uncertainties in computations involving charm quarks.  In
the sea sector, we employ CLS $N_{\rm f}=2+1$
configurations~\cite{Bruno:2014jqa}, with open boundary conditions in
the time direction~\cite{Luscher:2010iy, Luscher:2011kk,
  Luscher:2012av}, and fine values of the lattice spacing.
Concurrently, the use of a fully-twisted~\cite{Frezzotti:2000nk,
  Frezzotti:2004wz, Pena:2004gb} valence sector for the computation of
observables --- implying automatic ${\rm O}(a)$ improvement up to sea
quark mass effects --- along with the non-perturbative ${\rm O}(a)$
improvement of the sea action, is expected to result in excellent
scaling properties without the need to tune operator-dependent
improvement coefficients.

Here we will concentrate in the development of the mixed-action setup,
by identifying an efficient strategy for the matching of the sea and
valence sectors and by studying the scaling properties of basic light
quark observables. First results will be provided for CLS symmetric
point ensembles with degenerate quark masses,
$m_{\mathrm{u}}=m_{\mathrm{d}}=m_{\mathrm{s}}$. We refer to
Ref.~\cite{Collins:2017iud} for a study of charm quark observables
using CLS ensembles with the ${\rm O}(a)$ improved Wilson formulation.

\section{Lattice Formulation}\label{sec-1}

\subsection{Sea Sector}\label{sea_q}

The set of gauge ensembles used in our study were produced within the
CLS initiative~\cite{Bruno:2014jqa}. The lattice action involves the
Lüscher-Weisz tree-level improved gauge action~\cite{Luscher:1984xn},
\begin{equation} 
  S_g[U]=\frac{1}{g_0^2} \left( c_0\, \sum_{p} 
  \mathrm{tr} \left\lbrace 1-U(p) \right\rbrace + c_1 \sum_{r} \mathrm{tr} \left\lbrace 1-U(r) \right\rbrace \right)
  \, ,
  \label{act_gauge}
\end{equation}
where $p$ and $r$ run over the plaquettes and rectangles of the
lattice and their respective coefficients are $c_0 = 5/3$ and
$c_1=-1/12$.

The fermionic action comprises a Wilson Dirac
operator~\cite{Wilson:1974sk} for $N_\mathrm{f}=2+1$ flavours,
including the Sheikholeslami–Wohlert
term~\cite{Sheikholeslami:1985ij},
\begin{equation}
  S_\mathrm{f}[U,\overline \psi,\psi]=
  a^4 \sum_{\mathrm{f}=1}^3 \sum_x\, \overline \psi_\mathrm{f}(x)\, \left[
    \frac{1}{2}\sum_{\mu=0}^3 \{\gamma_\mu(\nabla^*_\mu+\nabla_\mu) -a\nabla^*_\mu\nabla_\mu\}
    + \frac{i}{4} a c_{\mathrm{SW}} \sum_{\mu,\nu=0}^3 \sigma_{\mu\nu}\widehat F_{\mu\nu}+ m_0
    \right] \, \psi_\mathrm{f}(x)\, ,
  \label{act_ferm}
\end{equation}
where $\nabla_\mu$ and $\nabla^*_\mu$ are the forward and backward
covariant derivatives and a non-perturbative determination of the
coefficient $c_{\mathrm{SW}}$ is used~\cite{Bulava:2013cta}.

The present analysis is based on four values of the lattice spacing
ranging from 0.087 $\mathrm{fm}$ to 0.050 $\mathrm{fm}$, as shown in
Table~\ref{tab_ens}.  The ensembles lie along lines of constant trace
of the bare quark mass matrix $M_\mathrm{q}$,
\begin{equation}
  \mathrm{tr} M_\mathrm{q}=2m_{q,\mathrm{u}}+m_{q,\mathrm{s}} = \mathrm{const}\, ,
  \label{tr_mb}
\end{equation}
where $m_{\mathrm{q,f}} = m_{0,\mathrm{f}}-m_\mathrm{cr}$. In this
way, cutoff effects of ${\rm O}( a \mathrm{tr} M_\mathrm{q} )$,
appearing, in particular, in the Symanzik expansion of the bare
coupling, can be kept constant when changing the sea quark masses.

In the approach towards the physical point, it is in practice
beneficial to consider a renormalised chiral trajectory in terms pion
and Kaon masses that is related at lowest order in chiral perturbation
theory to the condition in eq.~(\ref{tr_mb}) in terms of bare quark
masses. Specifically, the following dimensionless quantities,
\begin{equation}
  \phi_2=8t_0\,m_\pi^2\,, \qquad \qquad  \phi_4=8t_0\left( m_K^2 + \frac{1}{2}m_\pi^2 \right) \,,
  \label{phi}
\end{equation}
depending on the gradient flow observable $t_0$ defined in
section~\ref{Section: observables}, are used to establish the chiral
trajectory~\cite{Bruno:2014jqa, Bruno:2016plf}. Further details about
the choice of the chiral trajectory will be given in section~\ref{md}.
In addition to the chiral trajectories targeting the physical point,
further ensembles with mass degenerate $u, d$ and $s$ quarks ---
i.e. along a symmetric line that approaches the three flavour massless
theory --- are also available for $\beta=3.46$, as indicated in
Table~\ref{tab_ens}.

The simulations use periodic boundary conditions in the spatial
directions and open boundary conditions (OBC) in time. The efficiency
of the HMC algorithm to sample the topological sectors rapidly
deteriorates when the lattice spacing is reduced below $0.05$ fm. The
use of OBC allows to connect the field space in the continuum theory
by letting the topological charge to flow smoothly in and out of the
bulk of the lattice through the boundaries~\cite{Luscher:2010iy,
  Luscher:2011kk, Luscher:2012av}. As a result, the scaling law of the
autocorrelation times is expected to improve significantly. In this
way, the results from fine values of the lattice spacing --- which are
instrumental to control lattice artefacts proportional to the heavy
quark mass --- should also benefit from a reliable estimate of the
statistical uncertainties.
\begin{table}[tb]
  \begin{center}
    \small
    \begin{tabular}{cccccccc}
      \toprule
      Id &   $\beta$ &  $N_\mathrm{s}$  &  $N_\mathrm{t}$  & $m_\pi$[MeV] &   $m_K$[MeV] &  $m_\pi L$\\
      \midrule
      H101 & 3.40 & 32 & 96	& 420 &420  & 5.8\\
      H105 & 3.40 & 32 & 96	& 280 &460  & 3.9\\
      \midrule                                                   
      H400 & 3.46 & 32 & 96   & 420 & 420 & 5.2\\
      H401 & 3.46 & 32 & 96   & 550 & 550 & 7.3\\
      H402 & 3.46 & 32 & 96   & 450 & 450 & 5.7\\
      \midrule                                                   
      H200 & 3.55 & 32 & 96	& 420	&420  & 4.3\\
      N202 & 3.55 & 48 & 128	& 420	&420  & 6.5\\
      \midrule		                                               
      N300 & 3.70 & 48 & 128	& 420	&420  & 5.1\\
      \bottomrule
    \end{tabular}
    \caption{\label{tab_ens} List of CLS $N_\mathrm{f}=2+1$
      ensembles~\cite{Bruno:2014jqa} used in the present study. The
      values of the inverse bare coupling, $\beta=6/g^2_0$, correspond
      to the following approximate values of the lattice spacing,
      $a=0.087\,\mathrm{fm}$, $0.077\,\mathrm{fm}$,
      $0.065\,\mathrm{fm}$ and $0.050
      \,\mathrm{fm}$~\cite{Bruno:2016plf}. In the third and fourth
      columns, $N_\mathrm{s}$ and $N_\mathrm{t}$, refer to the spatial
      and temporal extent of the lattice. Approximate values of the
      pion and Kaon masses are provided. }
  \end{center} 
\end{table}

\subsection{Valence Sector}\label{val_q}

The mixed action uses in the valence sector a Wilson twisted mass
Dirac operator~\cite{Frezzotti:2000nk,Frezzotti:2004wz, Pena:2004gb,
  Sint:2007ug, Shindler:2007vp} where, in addition to a {\it standard}
mass term and to the chirally rotated mass term, the
Sheikholeslami–Wohlert term is also included,
\begin{equation}
  \frac{1}{2}\sum_{\mu=0}^3 \{\gamma_\mu(\nabla^*_\mu+\nabla_\mu) -a\nabla^*_\mu\nabla_\mu\}
  + \frac{i}{4} a c_{\mathrm{SW}} \sum_{\mu,\nu=0}^3 \sigma_{\mu\nu}\widehat F_{\mu\nu}+ \bm{m}_0 + i \gamma_5\, \bm{ \mu_0 }\,.
\end{equation}
This choice of the lattice action enforces that the renormalisation
factors of sea and valence sectors coincide.

Maximal twist is achieved once the bare {\it standard} mass is set to
the critical mass, $\bm{m}_0 = m_\mathrm{cr} \mathds{1}$, while the
twisted mass, $\bm{ \mu_0 }=\text{diag} \left( \mu_{0,\mathrm{u}},
\mu_{0,\mathrm{d}}, \mu_{0,\mathrm{s}}, \mu_{0,\mathrm{c}} \right)$ is
associated to the physical bare quark masses.  We note that this
target physical quark mass can be achieved by using both signs of the
twisted mass parameter $\mu_{0,\mathrm{q}}$.  At maximal twist, the
symmetries of the valence action guarantee the absence of ${\rm O}(a)$
lattice artefacts proportional to a hadronic scale $\Lambda$ and to
the valence quark masses
$\mu_\mathrm{0,\mathrm{q}}$~\cite{Frezzotti:2004wz}.  While residual
lattice artefacts of ${\rm O}(a g_0^4 \mathrm{tr} M_\mathrm{q})$ ---
arising from the sea quark mass matrix $M_\mathrm{q}$ --- can appear,
this mixed action setup should profit in the heavy quark sector from
the absence of the leading lattice artefacts proportional to the heavy
quark masses.

\subsection{Matching of Sea and Valence Quark Masses}\label{matching}

As mentioned in the previous section, at maximal twist, the {\it
standard} quark mass has to be set to its critical value in order to
enforce that its renormalised value vanishes in continuum limit. This
can for instance be achieved by imposing that a given observable
preserves a continuum symmetry at finite values of the lattice
spacing. More specifically, the twisted vector symmetry can be
restored by tuning to zero the valence PCAC quark mass,
$\left. m_{12}^{\mathrm{R}} \right|_\mathrm{v}$, in the light-quark
sector.\,\footnote{In what follows, the notation ``$|_\mathrm{v}$''
denotes the valence sector while ``$|_\mathrm{s}$'' refers to the
sea sector. For quark masses, the subscripts $1$ and $2$ refer to
two distinct light-quark flavours with degenerate masses, $m_1=m_2$,
and the presence of a superscript ${\rm R}$ denotes renormalised
quantities.}

In practice, this tuning process turned out to be straightforward
since, for all the considered ensembles, a short linear interpolation
in $\left. a m_{0,\mathrm{u}} \right|_\mathrm{v} =
1/\left(2\kappa_{\mathrm{u}}\rvert_\mathrm{v}\right) - 4$ around the
vanishing PCAC quark mass, $\left. m_{12}^\mathrm{R}
\right|_\mathrm{v}=0$, was sufficient to fulfil the maximal twist
condition.

In order to recover the unitarity of the theory in the continuum
limit, it is essential to match the valence and sea quark masses on
all the ensembles.  This can be achieved by enforcing that the
renormalised twisted mass is equal to the renormalised PCAC quark mass
of the sea, $\left. \mu_{1}^\mathrm{R} \right|_\mathrm{v} \; \equiv \;
\left. m_{12}^\mathrm{R} \right|_\mathrm{s}$.  Including ${\rm O}(a)$
counterterms in the evaluation of $\left. m_{12}^\mathrm{R}
\right|_\mathrm{s}$~\cite{Bhattacharya:2005rb}, this matching
condition reads,
\begin{equation}
  \label{eq:matching}
  \frac{1}{Z_\mathrm{P}}\, \mu_1 \equiv
  \frac{Z_\mathrm{A}}{Z_\mathrm{P}}\, \left. m_{12} \right|_\mathrm{s}
  \left( 1 + \left( \tilde{b}_\mathrm{A} - \tilde{b}_\mathrm{P}
  \right) a \left. m_{12} \right|_\mathrm{s} + \left(
  \overline{b}_\mathrm{A} - \overline{b}_\mathrm{P} \right) a\,
  \mathrm{tr} \left. M_\mathrm{q} \right|_\mathrm{s} \right)\, ,
\end{equation}
where the $Z_\mathrm{P}$ renormalisation factors appearing on both
sides of the equation are equal and can therefore be dropped, while
the $Z_\mathrm{A}$ factor has been determined non-perturbatively in
Ref.~\cite{DallaBrida:2017}. The bare sea quark mass $m_{12}$ includes
the ${\rm O}(a)$ improvement term proportional to $c_\mathrm{A}$,
which has been computed non-perturbatively in
Ref.~\cite{Bulava:2015bxa}. The mass-dependent $b$-type improvement
coefficients appearing in eq.~(\ref{eq:matching}) have been computed
to one-loop in perturbation theory~\cite{Taniguchi:1998pf} while
recent work to determine them non-perturbatively has been reported in
Refs.~\cite{deDivitiis:2017vvw, Korcyl:2016ugy}.

Alternatively, the quark mass matching can be done by tuning the sea
and valence pion masses to be equal, $\left. m_\pi \right|_\mathrm{v}
\; \equiv \; \left. m_\pi \right|_\mathrm{s}$. In this approach, both
side sides of the equation are automatically ${\rm O}(a)$ improved.

\section{Observables}\label{Section: OBC}

The presence of open boundary conditions in the Euclidean time
direction modifies the theory in the neighbourhood of the
boundaries. The ground state in the boundary $\ket{\Omega}$ is
different from the ground state in the bulk $\ket{0}$, which
asymptotically corresponds to the vacuum of the theory. In the
transfer matrix formalism, a typical two-point function is given by,
\begin{equation}
  \braket{X(x)P(y)}=\frac{1}{\mathcal{Z}}\braket{\Omega|e^{-(T-x_0)\hat{H}}X(\bm{x})e^{-(x_0-y_0)\hat{H}}P(\bm{y})e^{-y_0\hat{H}}|\Omega}\,,
\end{equation}
where the partition function is, $ \mathcal{Z}=
\braket{\Omega|e^{-T\hat{H}}|\Omega}$, and $\hat{H}$ stands for the
Hamiltonian of the system.

In order to address non-trivial boundary effects induced by the open
boundary conditions, fermionic observables are extracted through
appropriate ratios that cancel these effects. Examples of such ratios
will be provided in section ~\ref{Section: observables}.

\subsection{Mass Corrections}\label{md}

The renormalised chiral trajectory is chosen to correspond to a fixed
value of $\phi_4$, in eq.~(\ref{phi}), that passes through the
physical point. This typically requires the application of small mass
corrections to the simulated bare quark masses, that were set via
eq.~(\ref{tr_mb}) in the dynamical simulations. This small variation
of the bare quark masses can be applied to any observable through a
low order Taylor expansion~\cite{Bruno:2016plf}.  For a generic
derived observable defined as a function $f\left( \bar{A}_1, \dots,
\bar{A}_n \right)$ of the mean values of primary observables
$\bar{A}_i=\left\langle A_i \right\rangle$, the expansion with respect
to a shift of the mass quarks reads,
\begin{equation}
  f(m'_{\mathrm{q,u}},m'_{\mathrm{q,s}})=f(m_{\mathrm{q,u}},m_{\mathrm{q,s}})+2(m'_{\mathrm{q,u}}-m_{\mathrm{q,u}}) \frac{df(m_{\mathrm{q,u}},m_{\mathrm{q,s}})}{dm_{\mathrm{q,u}}} + 
  (m'_{\mathrm{q,s}}-m_{\mathrm{q,s}}) \frac{df(m_{\mathrm{q,u}},m_{\mathrm{q,s}})}{dm_{\mathrm{q,s}}} ,
\end{equation}
where the derivative terms read,
\begin{equation}
  \frac{df}{dm_\mathrm{q}} = \sum_i \frac{\partial f}{\partial \bar{A}_i} \left[ 
    \left\langle \frac{\partial A_i}{\partial m_\mathrm{q}} \right\rangle
    - \left\langle \left( A_i -  \bar{A_i} \right)
    \left( \frac{\partial S}{\partial m_\mathrm{q}} - \frac{\overline{\partial S}}{\partial m_\mathrm{q}} \right) \right\rangle
    \right] ,
\end{equation}
and $S$ stands for the action.

Notice that for Wilson twisted mass (Wtm) valence fermions only terms
proportional to derivative of the action can contribute to the
expansion.

\subsection{Computation of Observables}\label{Section: observables}

We use the previously discussed mixed action approach to compute light
quark observables involved in a continuum limit scaling study of our
lattice formulation. The computation of sea sector observables is
relevant for the matching procedure and to compare relative cutoff
effects between sea and valence sectors.

We employ the gradient flow quantity $t_0$ as a relative reference
scale. It is defined via the dimensionless observable $t^2
\braket{E(t,x_0)}$ \cite{Luscher:2010iy, Luscher:2010we}, where the
energy density $E(t,x_0)$ is evaluated at a fixed value of the flow
time $t$,
\begin{equation}
  t^2 \braket{E^{\mathrm{av}}(t)}\rvert_{t=t_0}=0.3\,.	
\end{equation}
The superscript ``$\mathrm{av}$'' stands for the Euclidean time
average over the plateau range in $x_0$.

Pseudoscalar meson masses and decay constants as well as PCAC quark
masses are extracted from the two-point correlation functions,
\begin{equation}
  \begin{split}
    f_\mathrm{P}(x_0,y_0)&= a^6\sum_{\vec{x},\vec{y}}\braket{P(x)P(y)}
    \,,\\
    f_\mathrm{A}(x_0,y_0)&= a^6\sum_{\vec{x},\vec{y}}\braket{A_0(x)P(y)}\,,
  \end{split}
\end{equation}
where $P$ and $A_0$ are the pseudoscalar density and the time
component of the improved axial current,
$A_0=\bar{\psi}\gamma_0\gamma_5\psi+ac_A\partial_0
\bar{\psi}\gamma_5\psi$, respectively. We set the source of the
two-point correlation function close to each boundary and symmetrise
with respect to $x_0$ by taking advantage of the time reversal
symmetry.

Pseudoscalar masses are extracted by averaging over a plateau range the effective mass,
\begin{equation}
  a m_\mathrm{eff}(x_0)= \log \frac{f_\mathrm{P}(x_0,a)}{f_\mathrm{P}(x_0+a,a)}\,.
\end{equation}

The determination of the light PCAC quark masses $m_{12}$ is required
for the matching procedure. We determine the PCAC masses through the
following expression,
\begin{equation}
  m_{12}=\left(\frac{\tilde{\partial}_{0} f_\mathrm{A}(x_0,a)}{2
    f_\mathrm{P}(x_0,a)}\right)^{\mathrm{av}}\,,
\end{equation}
where $\tilde{\partial}_0$ stands for the symmetric definition of the
numerical derivative with respect to the Euclidean time.

Decay constants can be obtained through the Euclidean time average of
a ratio,
\begin{equation}
  \label{Eq: Ratio}
  R_X(x_0)=\sqrt{\frac{f_X(x_0,a) \,f_X(x_0,T-a)}{f_\mathrm{P}(T-a,a)}}\,,
\end{equation}
where $X$ is a fermionic operator. The ratio $R_X(x_0)$ is constructed
in order to cancel boundary effects and to isolate the desired matrix
elements from a fit to the plateau region. Specifically, the
pseudoscalar meson decay constant of the valence sector can be
computed in the following way,
\begin{equation}
  f_{\mathrm{PS}}\rvert_\mathrm{v}=\left(\frac{2}{m_{PS}L^3}\right)^{1/2}R^{\mathrm{av}}_P\,.
\end{equation}
By exploiting the PCVC relation, the valence decay constant can be
obtained without an explicit use of renormalisation constants.

Isospin breaking effects induced by the twisted mass regularisation
can be monitored through the difference between charged pion
$m_{\pi^\pm}$ and neutral pion $m_{\pi^0}$ masses. In our mixed action
approach, these effects are confined to the valence sector and,
therefore, the neutral connected pion mass $m_{\pi^{(0,\mathrm{c})}}$
already provides a good probe to analyse the size of these $O(a^2)$
lattice artefacts. At leading order in Wilson chiral perturbation
theory, this mass splitting can be related to the low energy constant
$w'_8$~\cite{Hansen:2011kk}, in the following way,
\begin{equation}
  m^2_{\pi^\pm}-m^2_{\pi^{(0,c)}}=a^2\,w'_8\,.
\end{equation}

For various observables considered in our study, a criterion for
choosing the Euclidean time plateaux is required in order to remove
boundary effects and excited state contributions. Following the
proposal of Refs.~\cite{oai:export, Bruno:2016plf}, we define the
plateau regions by imposing that statistical errors are at least four
times larger than the systematic errors from the first excited state
at the starting point of the plateau,
$\delta_{\mathrm{stat}}(x_{0,\mathrm{min}})\gtrsim
4\delta_{\mathrm{syst}}(x_{0,\mathrm{min}})$.  For the ensembles
considered in this study, it turned out that for pseudoscalar mesons
the excited state contamination can be neglected for $x_0\gtrsim
4\sqrt{8t_0}$ in the Wilson case, whereas $x_0\gtrsim 5\sqrt{8t_0}$ is
needed with Wilson twisted mass fermions.

\section{Numerical Studies}

\subsection{Tuning to Maximal Twist}

\begin{figure}
  \begin{center}
    \subcapraggedrighttrue
    \subfigure[\label{H105r005_pcac_tun}]{
      \includegraphics[scale=0.38]{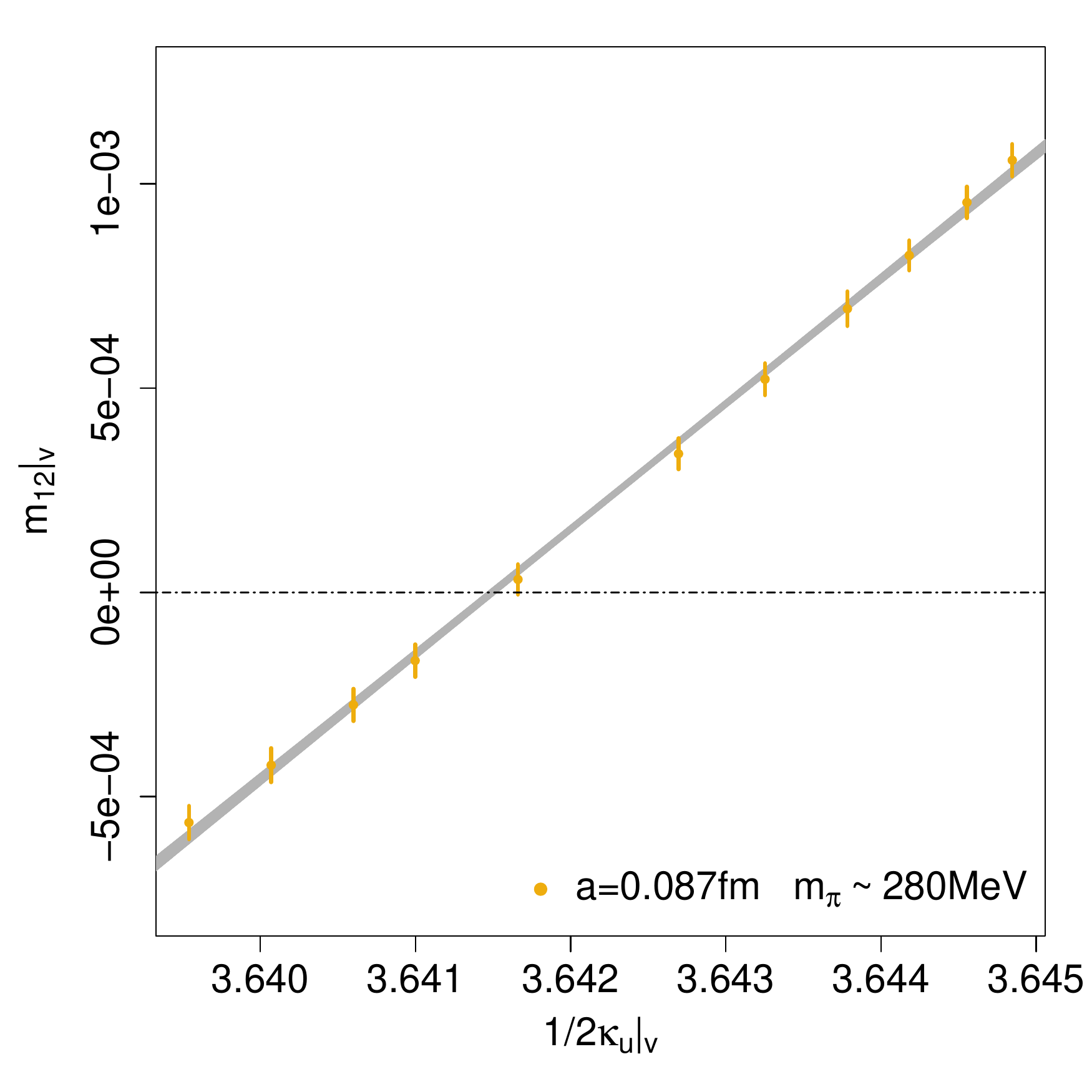}%
    }
    \quad
    \subfigure[\label{H105r005_meson_tun}]{
      \includegraphics[scale=0.38]{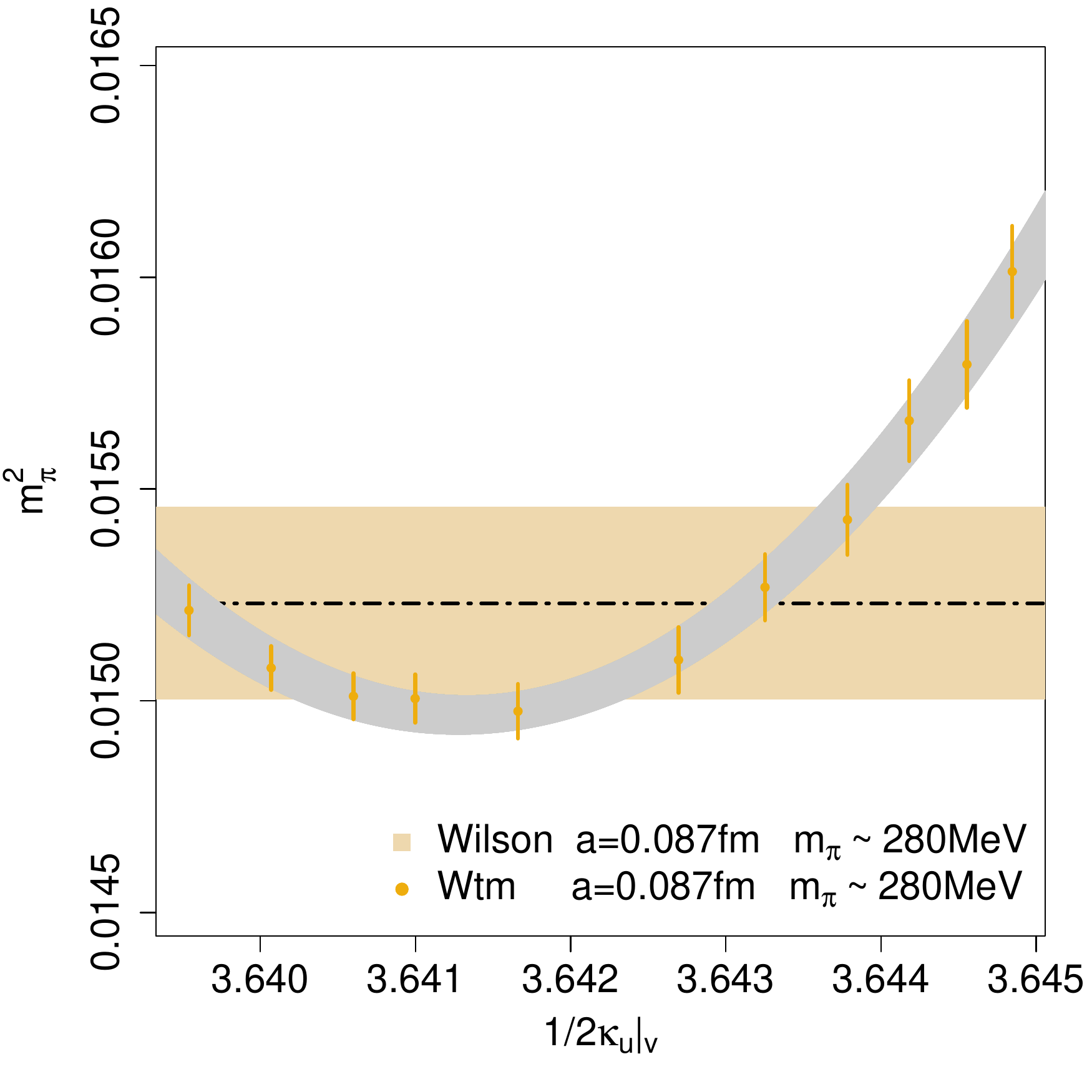}%
    }
    \caption{(a) Tuning to maximal twist by linearly interpolating to
      zero the valence PCAC quark mass as a function of
      $\left. 1/2\kappa_u\right|_\mathrm{v}$. (b) Dependence of the
      pion mass squared on $\left. 1/2\kappa_u\right|_\mathrm{v}$
      around maximal twist. Data points show the expected parabolic
      behaviour of $\left. m_\pi^2\right|_\mathrm{v}$ in the
      neighbourhood of maximal twist. The horizontal yellow band shows
      the corresponding result from the Wilson formulation. In both
      panels, the data points correspond to the ensemble H105 with
      coarsest lattice spacing (see Table~\ref{tab_ens}). The sea and
      valence quark masses were matched by imposing,
      $\left. \mu_{1}^\mathrm{R} \right|_\mathrm{v} \, \equiv \,
      \left. m_{12}^\mathrm{R} \right|_\mathrm{s}$, as explained in
      section~\ref{matching}.}
    \label{fig_H105r005}
  \end{center}
\end{figure}

The tuning to maximal twist can be performed by a linear interpolation
to a vanishing valence PCAC quark mass, $\left. m_{12}
\right|_\mathrm{v}$, as a function of $\left.\kappa_\mathrm{u}
\right|_\mathrm{v}^{-1}$. This interpolation is illustrated in
Figure~\ref{H105r005_pcac_tun}.  The tuning can thus be achieved
through a few simulation points in the neighbourhood of
$\left. m_{12}^\mathrm{R} \right|_\mathrm{v}=0$. At lowest order in
chiral perturbation theory, the pion mass squared is given by,
\begin{equation}
  m_\pi^2 \propto \sqrt{ \left(m^\mathrm{R}\right)^2 +
    \left(\mu^\mathrm{R}\right)^2 } \,,
\end{equation}
where $m^\mathrm{R}$ and $\mu^\mathrm{R}$ are the standard and twisted
renormalised quark masses, respectively. For fixed values of
$\mu^\mathrm{R}$, $m_\pi^2$ is expected to follow a parabolic
behaviour in $m^\mathrm{R}$ in the neighbourhood of its minimum,
$m^\mathrm{R}=0$.  Figure~\ref{H105r005_meson_tun} shows an example of
such a parabolic behaviour around maximal twist.  The horizontal band
provides a comparison to the measurement of the pion mass through the
Wilson regularisation when using the same statistics as in the Wtm
case.  Although relative cutoff effects of ${\rm O}(a^2)$ could be non
negligible at this coarsest value of the lattice spacing, we observe
that the sea and valence quark masses agree within the errors.  The
fact that the minimum of the parabola coincides with the vanishing of
the PCAC mass in Figure~\ref{H105r005_pcac_tun} provides a cross-check
of the identification of the maximal twist point.

As indicated in Section~\ref{Section: OBC} and in Table~\ref{tab_ens},
the three ensembles at $\beta=3.46$ lie along the symmetric line which
crosses the three flavour massless theory point. By extrapolating our
determination of the bare standard mass at maximal twist along the
symmetric line, we obtain an estimate of the critical mass,
$\kappa_\mathrm{cr}$, which is in the same ballpark as an independent
determination coming from the Schr\"odinger Functional
scheme~\cite{Fritzsch:2017} (see also Ref.~\cite{Bali:2016umi} for a
study using large volume CLS ensembles).

\subsection{Continuum Limit Scaling of $m_\pi$ and $f_\pi$}

We perform a continuum-limit scaling study of the light pseudoscalar
meson decay constant $f_\pi$ in order to check the universality of our
mixed action setup by a direct comparison of the continuum results of
the sea and valence formulations. Furthermore, this study provides an
indication of the relevance of residual $O(a)$ cutoff effects
proportional to sea quark masses. The continuum limit scaling is
carried out along a line of constant physics composed of symmetric
point ensembles, $m_{\mathrm{u}}=m_{\mathrm{s}}$, for which
$\phi_4\rvert_{\mathrm{s}} = 1.11$~\cite{Bruno:2016plf}.

\begin{figure}
  \centering
  \includegraphics[scale=1.0]{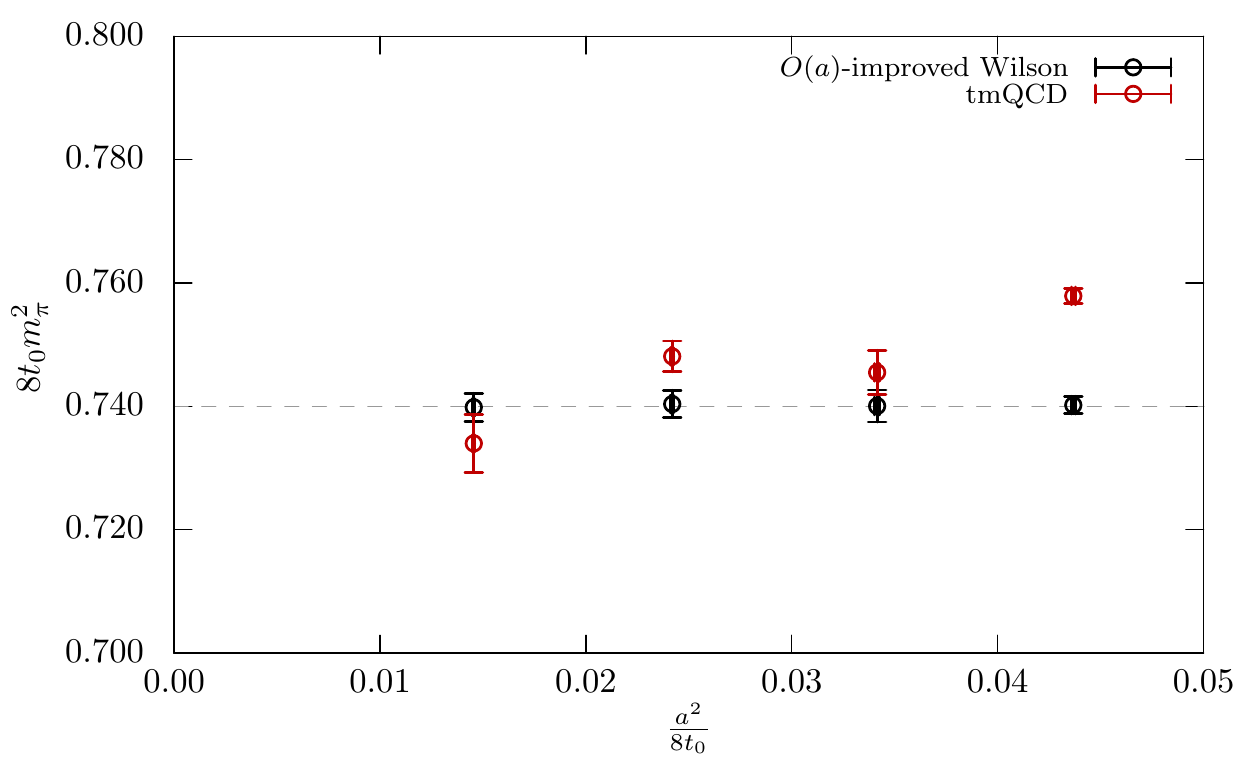}
  \caption{Continuum limit scaling of $m_\pi^2$ in units of $t_0$ for
    various symmetric point ensembles,
    $m_\mathrm{u}=m_\mathrm{d}=m_\mathrm{s}$. Black points refer to
    the improved Wilson fermions calculations of
    Ref.~\cite{Bruno:2016plf} where the ensembles were tuned to
    $\phi_4=1.11$, as shown by the grey discontinuous line. Red points
    are preliminary determinations from our fully twisted tmQCD mixed
    action setup. The renormalised light quark mass in the valence and
    in the sea were matched such that, $\left. \mu_{1}^\mathrm{R}
    \right|_\mathrm{v} \; \equiv \; \left. m_{12}^\mathrm{R}
    \right|_\mathrm{s}$.}  \captionsetup{justification=centering}
  \label{mpi_scaling}
\end{figure}

\begin{figure}
  \centering
  \includegraphics[scale=1.0]{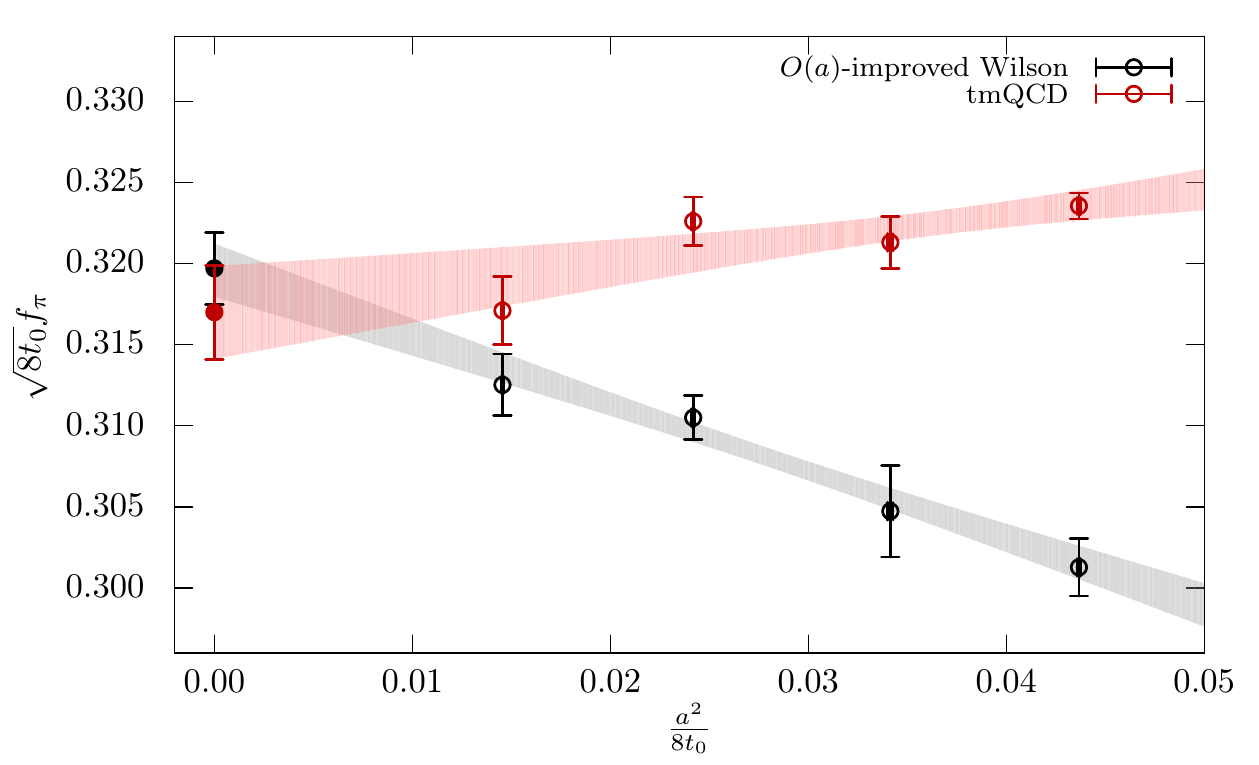}
  \caption{Continuum limit scaling of the pion decay constant $f_\pi$
    in units of $t_0$ for various symmetric point ensembles,
    $m_\mathrm{u}=m_\mathrm{d}=m_\mathrm{s}$. Black points refer to
    the improved Wilson fermions calculations of
    Ref.~\cite{Bruno:2016plf}. Red points are preliminary
    determinations from our fully twisted tmQCD mixed action
    setup. The matching between sea and valence has been performed by
    imposing $\left. \mu_{1}^\mathrm{R} \right|_\mathrm{v} \; \equiv
    \; \left. m_{12}^\mathrm{R} \right|_\mathrm{s}$.}
  \captionsetup{justification=centering}
  \label{fpi_scaling}
\end{figure}

As mentioned in section~\ref{matching}, the matching can been
performed by imposing that renormalised quark masses of sea and
valence quark masses are equal. As a result, relative cutoff effects
of $O(a^2)$ are expected between the pseudoscalar meson masses of the
sea and valence sectors. From Fig.~\ref{mpi_scaling}, we observe that
these relative cutoff effects are negligible for the lattice spacings,
$a=0.077\,\mathrm{fm}$ and $a=0.050\,\mathrm{fm}$, while they
approximately amount to a 1.5\% and a 3\% effect at
$a=0.065\,\mathrm{fm}$ and $a=0.087\,\mathrm{fm}$, respectively. An
ongoing complementary study, where the matching of sea and valence
quark masses is done through the pion masses, will allow to monitor
the continuum scaling of the light quark masses.
\newline
In Fig.~\ref{fpi_scaling}, we present preliminary results for the
continuum limit scaling of the pion decay decay constant $f_\pi$ in
the the sea~\cite{Bruno:2016plf} and valence sectors. The scaling
behaviour is consistent with $O(a)$ improvement for both sea and
valence regularisations. Furthermore, the expected agreement of the
continuum limit results is verified.

\subsection{Pion Mass Splitting}

\begin{figure}
  \centering
  \includegraphics[scale=1.0]{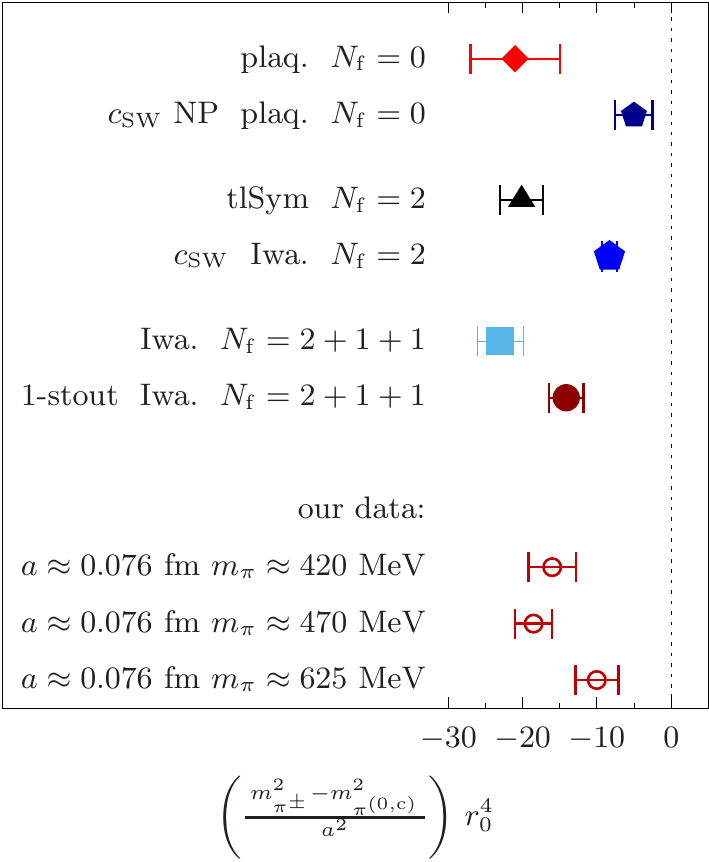}
  \caption{Comparison of the mass splitting between the charged and
    the neutral connected pions for various lattice
    actions~\cite{Herdoiza:2013sla, Abdel-Rehim:2015pwa} in units of
    the Sommer scale $r_0$. The quantity in the horizontal axis is
    related to the low energy constant $w'_8$ of Wilson chiral
    perturbation theory, parametrising the scale of $O(a^2)$ lattice
    artefacts in this pion mass splitting.  }
  \captionsetup{justification=centering}
  \label{mpic_mpi0c}
\end{figure}

In Fig.~\ref{mpic_mpi0c}, we compare our preliminary determination of
the mass splitting between the charged pion and the connected
contribution to the neutral pion with results from different lattice
actions ~\cite{Herdoiza:2013sla, Abdel-Rehim:2015pwa}.  The mass
difference, $(m^2_{\pi^\pm}-m^2_{\pi^\mathrm{(0,c)}})\,r^4_0/a^2$, in
terms of the Sommer scale $r_0$, quantifies the scale of the $O(a^2)$
isospin breaking effects. We observe that our estimates of this mass
splitting fall on the same ballpark as those from other lattice
formulations ~\cite{Herdoiza:2013sla, Abdel-Rehim:2015pwa}.

\section{Conclusions and Outlook}

We define a regularisation based on a mixed action approach, aimed to
improve the control of systematic uncertainties in the charm quark
sector.  The setup combines a fully-twisted valence action matched
through the quark masses with the $\mathrm{O}(a)$ improved Wilson sea
fermionic action. The matching procedure is simplified since the sea
and valence actions share the same renormalisation factors. The tuning
to maximal twist is performed by linearly interpolating the valence
PCAC quark mass to the vanishing point. A first study of the continuum
limit scaling of the pion decay constant has shown the agreement of
the continuum results with respect to the Wilson formulation. This
provides an important validation of this mixed action formulation.

We plan to extend the present study to include an alternative matching
condition of sea and valence quark masses through the use of pseudoscalar
meson masses in order to profit from their automatic $O(a)$ improvement.
The next phase of the project will concern the extension of this mixed
action setup to the strange and charm sectors.

\subsection*{Acknowledgements}

We are grateful to CLS members for producing the gauge configuration
ensembles used in this study.  We wish to thank Mattia Bruno, Patrick
Fritzsch, Tomasz Korzec and Stefan Schaefer for useful discussions and
for providing valuable input for our analyses.
We thankfully acknowledge the computer resources at FinisTerrae II,
MareNostrum and Calendula and the technical support provided by CESGA,
BSC and FCSCL (FI-2016-3-0022, FI-2017-1-0041, FI-2017-2-0029).
We acknowledge support through the Spanish MINECO project
FPA2015-68541-P, the Centro de Excelencia Severo Ochoa Programme
SEV-2016-0597 and the Ramón y Cajal Programme RYC-2012-10819.

\nocite{*}
\newcommand{\noop}[1]{}

\end{document}